\documentclass[12pt,preprint]{aastex}

\begin{document}

\title{Helicity Observation of Weak and Strong Fields}

\author{Mei Zhang\altaffilmark{1}}

\altaffiltext{1}{National Astronomical Observatory,
Chinese Academy of Sciences, A20 Datun Road,
Chaoyang District, Beijing 100012, China; Email: zhangmei@bao.ac.cn}


\begin{abstract}
We report in this letter our analysis of a large
sample of photospheric vector magnetic field measurements.
Our sample consists of 17200 vector magnetograms obtained from 
January 1997 to August 2004 by Huairou Solar Observing Station
of the Chinese National Astronomical Observatory. Two physical
quantities, $\alpha$ and current helicity,
are calculated and their signs and amplitudes are studied 
in a search for solar cycle variations. Different from other 
studies of the same type, we calculate these quantities
for weak ($100G<|B_z|<500G$) and strong ($|B_z|>1000G$) fields
separately. For weak fields, we find that the signs of both
$\alpha$ and current helicity are consistent with the
established hemispheric rule during most years of the
solar cycle and their magnitudes show a rough tendency of
decreasing with the development of solar cycle.
Analysis of strong fields gives an interesting result:
Both $\alpha$ and current helicity present a sign
opposite to that of weak fields. Implications of these
observations on dynamo theory and helicity production
are also briefly discussed.
\end{abstract}

\keywords{MHD --- Sun: magnetic fields --- Sun: interior}


\section{Introduction}

Magnetic helicity is a physical quantity that
measures the topological complexity of a magnetic field, 
such as the degree of linkage and/or twistedness in the 
field (Moffatt 1985, Berger \& Field 1984). It has been shown 
that its total amount is approximately conserved in the Sun
even when there is an energy release during fast magnetic
reconnection (Berger 1984). This conservation of total
magnetic helicity is considered to play an important role 
in the dynamical processes in the Sun.
For example, by considering helicity conservation in the 
mean-field dynamo, theories have predicted that
solar dynamo would produce opposite helicity signs
in the mean field and in the fluctuations (Blackman 
\& Field 2000, see Ossendrijver 2003 for a review).
It has also been considered that magnetic helicity and its
conservation may play an important role in CME dynamics
(Low 2001, Demoulin et al. 2002) where accumulation of
total magnetic helicity in the respective northern and 
southern hemispheres leads to a natural magnetic energy 
storage for CME eruptions (Zhang \& Low 2005, Zhang, Flyer
\& Low 2006).

A direct measurement of magnetic helicity and hence a direct test
of above theories by observations are still out of our reach
because so far the photosphere is still the only layer that we can
measure vector magnetic fields with reasonable temporal and
spatial resolutions. However, by calculating derived physical
quantities, such as $\alpha$ and current helicity, from 
observed photospheric vector magnetograms we do get a glimpse 
of properties of magnetic helicity in the Sun. For example,
from photospheric magnetic field measurements we learn that 
magnetic fields emerging from the solar convection zone to the
photosphere are already significantly twisted (Kurokawa 1987,
Leka et al. 1996) and statistically these fields possess 
a positive helicity sign in the southern hemisphere and 
a negative helicity sign in the northern hemisphere 
(Pevtsov et al. 1995, Bao \& Zhang 1998).
These observations thus provide us implications on how
magnetic helicity might be produced in the convection zone
(Berger \& Ruzmaikin 2000) and how magnetic helicity conservation
might have played a role in balancing the twist and writhe 
helicity in an originally untwisted flux rope (Longcope et al. 1998).

In this letter, we intend to use photospheric vector
magnetic field measurements to find further observational
indications of helicity production and conservation.
Different from other works of the same type, we separate 
studied fields into two parts: strong magnetic fields and 
weak magnetic fields.
We organize our paper as follows: In \S 2, we describe our
observation and data reduction. In \S 3, we present our 
analysis and discussions. We conclude the letter with 
a brief summary in \S 4.


\section{Observation and Data Reduction}

The tunable birefringent filter of the solar telescope
magnetograph at the Huairou Solar Observing Station of the
Chinese National Astronomical Observatory can be aimed at 
different passbands for different observations 
(Ai \& Hu 1986). For photospheric observations
the passband of the filter is set in the FeI$\lambda$5324 
line: at 0.075\AA \hspace{1mm} from the line center for 
the measurement of longitudinal magnetic field (Stokes V) 
and at the line center for the measurement of transverse 
magnetic fields (Stokes Q and U). More information of the 
magnetograph and calibration can be found in Ai et al. 
(1982) and Zhang \& Ai (1986).

A dataset of photospheric vector magnetograms obtained
by above magnetograph during the period of 1997 January 1 
to 2004 August 31 is analyzed in this letter.
This dataset contains 17200 vector magnetograms and covers 
almost all active regions appeared during this period.
We calibrate each vector magnetogram according to Ai et al.
(1982) and solve the 180-degree ambiguity by setting 
the directions of transverse fields most closely to a
current-free field.

We calculate two physical quantities, $\alpha$ and current 
helicity, of each magnetogram, as helicity proxies.
We calculate $\alpha$, either as a best-fit single value 
$\alpha_{best}$ following Pevtsov et al. (1995) or as a mean 
value $\langle\alpha_z\rangle$ of the local
$\alpha_z=(\nabla\times{\bf B})_z / B_z$ 
as in Pevtsov et al. (1994).
The two $\alpha$ values so calculated are all indicators of 
the twistedness of the measured field and there is a linear 
relationship between them when derived from the same set of
magnetograms (Burnette et al. 2004).
We shall use $\alpha_{best}$ in \S 3.1 in comparison with
Pevtsov et al. (2001) and $\langle\alpha_z\rangle$ in \S 3.2 
and \S 3.3 for $\langle\alpha_z\rangle$ is presumably less 
dependent on the linear force-free assumption.
The current helicity is calculated as
$h_c=B_z\cdot(\nabla\times{\bf B})_z$, which is actually
the longitudinal ($z$) component of the current helicity
density at the photosphere ($z=0$). When calculating these 
quantities we have used only those magnetograms
whose longitudes are less than 40 degrees from the disk
center and only those data points whose longitudinal flux
densities ($|B_z|$), after the correction of projection 
effect, are greater than 100G and whose transverse flux 
densities ($|B_x|$ and $|B_y|$) are both greater than 200G.
Note our treatments in data reduction so far are as typical as
most other authors in reducing vector magnetograms 
(Pevtsov et al., 1994, 1995, 2001; Bao \& Zhang 1998).

Our unique treatment of the data is that we divide our studied
fields into two parts: strong magnetic fields whose longitudinal 
flux densities ($|B_z|$) are greater than 1000G, and weak
magnetic fields whose longitudinal flux densities ($|B_z|$)
are between 100G and 500G. By such a definition, our strong
fields are then mainly consisted of the umbra of sunspots 
and our weak fields of the enhanced magnetic networks around 
sunspots. We calculate $\alpha$ and current helicity
for such defined strong and weak fields separately. Note 
by doing so, not only we gain the opportunity to study the
possible differences between weak and strong fields within 
active regions, but also we get a chance to learn indicated
helicity properties of the global Sun if we identify our
observed weak fields as the representatives of the general
weak fields distributed over the whole surface.


\section{Analysis and Discussion}


\subsection{Comparison with previous studies}

Before we proceed to present our results it is useful to 
check our data reduction of this dataset with previous 
results obtained by other instruments and datasets.
We select a subsample of our dataset, containing observations
made between 1997 July to 2000 September, in order to compare
with Pevtsov et al. (2001) where $\alpha_{best}$ and current 
helicity were also calculated for the same period of time.
The difference is that their magnetograms were obtained by the
Haleakala Stokes Polarimeter (HSP) at Mees Solar Observatory.

Figure 1 presents the latitudinal profile of $\alpha_{best}$
for the 391 active regions observed by Huairou magnetograph
during this period of time. Each point presents the
average value of $\alpha_{best}$ when multiple magnetograms
of the same active region were obtained. Note in producing 
this figure we did not separate the weak and strong fields 
but instead use all data points with $|B_z|>100G$ and 
$|B_x, B_y|>200G$, in order to make a reasonable comparison 
with Pevtsov et al. (2001). The green line shows the 
least-square best-fit linear function of these $\alpha_{best}$
values. The similarity between our figure and Figure 1 of 
Pevtsov et al. (2001) indicates a good consistence between
the two datasets.

Out of our 391 active regions during this period, 58.9\% of
214 active regions in the northern hemisphere have
$\alpha_{best}<0$ and 67.2\% of 117 active regions in
the southern hemisphere have $\alpha_{best}>0$. These numbers
are consistent with the numbers of 62.9\% and 69.9\% for the
northern and southern hemispheres respectively in Pevtsov
et al. (2001).
Our data shows no tendency of hemispheric rule by current helicity.
44.4\% of 214 active regions in the northern hemisphere have
$h_c<0$ and 45.8\% of 117 active regions in the southern 
hemisphere have $h_c>0$. Note in Pevtsov et al. (2001)
a much weaker tendency is also found with numbers of 50\% 
and 57.5\% for their $h_c$ values in the northern and southern
hemispheres respectively. They contribute this difference
to Faraday rotation. But we suggest the difference is
largely (although possibly not all) because of a physical 
point which we will return to address below.

Averages of $\alpha_{best}$ for active regions observed in each 
10 degrees of solar latitudes are also plotted in Figure 1,
presented as red square symbols. The large error bars of these 
averages remind us that our established hemispherical rule is 
of a statistical result. Individual active regions may present
large deviations from the mean values. This is also true for 
other statistical results that we will present below.


\subsection{Helicity observation of weak fields}

Figure 2 presents our result of solar cycle variations of
$\alpha$ (top panel) and current helicity (middle panel)
for weak fields ($100G<|B_z|<500G$). Each point in these plots
is a weighted average of $\langle\alpha_z\rangle$ or current helicity
for active regions observed during one year. For active regions
in the southern hemisphere the weight is set to $1$ and for
active regions in the northern hemisphere the weight is set
to $-1$. The weighted averages then indicate the magnitudes of
$\alpha$ or current helicity averaged over the global
surface during a whole year, assuming the northern and
southern hemispheres have opposite helicity signs.
We see that both averaged $\alpha$ and current helicity
have positive signs except for the Year 2004. This tells us
that both $\alpha$ and current helicity for weak fields
obey the established hemispheric rule during most years of the
solar cycle. The averaged $\alpha$ and current helicity
for Year 2004 are negative, which indicates the usual hemispheric
rule is not followed in this year. This is consistent with 
Hagino \& Sakurai (2005) where they also found a violation of
the usual hemispheric rule during solar minimums.

Figure 2 also presents a rough tendency of a decrease of
$\alpha$ and current helicity with the development
of solar cycle. We notice in Berger \& Ruzmaikin (2000)
the helicity production rate by differential rotation in 
solar interior is calculated and their calculation also shows 
a similar decrease of magnitudes of the rate of helicity
transported into the northern and southern hemisphere respectively.
This can be seen from the bottom panel of Figure 2 where 
the helicity transportation rate into the southern hemisphere 
by the m=0 mode is replotted, with data taken from
Berger \& Ruzmaikin (2000). This interesting consistence
seems to suggest that differential rotation is the source
of helicity production in solar interior although we are not
able to make a conclusion because we do not know whether 
the $\alpha$ effect will also produce the same tendency or not.

As pointed out by the careful referee, the calculated transferred 
helicity ends at zero during solar minimums whereas our observation
as well as Hagino \& Sakurai (2005) show the helicity goes to 
the opposite sign during solar minimums. We intend to explain this
as a result of trans-equatorial reconnection (Pevtsov 2000)
which has consumed the helicity of the dominate sign in
each hemisphere, a point interesting of itself but is out of
the scope of current letter.

Another interesting implication of Figure 2 is that,
whereas we usually consider helicity variation as a function
of latitude as presented in Figure 1, another possibility is
that the helicity variation is more associated with
solar cycle dependence and the known latitude dependence is
just a derived relation from this solar cycle dependence of
helicity and the Butterfly diagram.


\subsection{Helicity observation of strong fields}

For strong magnetic fields ($|B_z|>1000G$), calculation of
weighted averages of $\alpha$ and current helicity
presents an interesting result, shown in Figure 3.
All averaged $\alpha$ and current
helicity are negative, which means they do not follow the
usual hemispheric rule. This also means that strong fields
have a helicity sign opposite to that of weak fields.

As we have mentioned earlier, if we interpret our observed 
weak fields in active regions as the representatives of the 
general weak fields distributed over the global Sun, then we 
may use them to represent the large-scale field. Our strong 
fields may be used to represent the small-scale fluctuations
compared to the large-scale of the global Sun. Then under this 
interpretation our observation seems to be consistent with the 
theory that solar dynamo would produce opposite helicity signs 
in the mean field and in the fluctuations.

It is also interesting to notice that in Berger \& Ruzmaikin 
(2000) the higher modes helicity, such as the m=5 mode 
replotted in Figure 3, also has a sign opposite to that
of the m=0 mode. Again, if we interpret their low-degree (such
as m=0) mode field corresponds to our weak field because
both of them represent a more uniformly-distributed field
over the global Sun and their high-degree (such as m=5) mode 
field corresponds to our strong field because both of them 
are sporadically appeared on the surface, then their 
calculation and our observation show a consistence again.

The observation that strong fields have a helicity sign 
opposite to that of weak fields may help us understand why
$\alpha_{best}$ usually shows a better hemispheric rule than
current helicity if both quantities are calculated from
vector magnetograms of the whole field (Pevtsov et al. 2001). 
We interpret it as follows. When we calculate $\alpha_{best}$
of the whole field, each data point is given an equal weight.
This results in the calculated $\alpha_{best}$ presenting
the sign of weak fields, whose number of data points dominates
over that of strong fields. But when we calculate the current 
helicity of the whole field, defined as
$h_c=B_z\cdot(\nabla\times{\bf B})_z = \alpha B_z^2$,
we have attributed a weight of $B_z^2$ to each data point.
This then results in a nearly cancellation of current helicity
between the weak and strong fields because weak and strong 
fields happen to have opposite helicity signs and the former 
has a larger number of data points but smaller $B_z^2$ values
for each data point whereas the latter has a smaller number of 
data points but each data point has a larger $B_z^2$ value.

It has been suggested that Faraday rotation contributes to
the difference between $\alpha_{best}$ and current helicity.
We suggest the main reason is the opposite helicity signs 
between weak and strong fields. J. T. Su \& H. Q. Zhang (2006,
in preparation) recently did a calculation and it shows
that whereas Faraday rotation may rotate the transverse fields
to 20 - 30 degrees, the resultant $\alpha$ values are less
influenced, with changes of $\alpha$ values all less than 
a few percentages. Another comment is that if Faraday rotation
is the reason of the difference we should not see the difference
in the dataset obtained by spectrograph-type magnetographs
where the effect of Faraday rotation can be taken care of by
inversion methods. But the difference is observed in Pevtsov
et al. (2001) where HSP data are used. We have recently
checked several active regions observed by ASP/HAO.
Similar feature of opposite helicity signs between weak and
strong fields is found, although not in every region
examined. Also kindly pointed out by the referee, similar
tendency of opposite helicity signs is also indicated in a 
decaying active region observed by ASP (Figure 4 of 
Pevtsov \& Canfield 1999).

Finally we point out another consistence of our observation
with previous study. By applying a known reconstruction
technique to MDI data Pevtsov and Latushko (2000) calculated
the current helicity of the global Sun. They found that the usual
hemispheric rule is followed for regions above 40 degrees of
solar latitudes whereas the rule is surprisingly not obvious
for regions within 40 degrees of solar latitudes.
With our observation, we now can interpret it as follows.
In high latitudes magnetic fields are dominated by weak fields
with their signs following the usual hemispherical rule, whereas
in low latitudes strong fields with an opposite helicity sign
present to result in a reduction to the usual hemispherical rule.


\section{Summary}

A large sample of 17200 photospheric vector magnetograms 
of active regions obtained from January 1997 to August 2004
is analyzed in this letter. Different from other works, 
we calculate the helicity proxies, $\alpha$ and current helicity, 
for weak ($100G<|B_z|<500G$) and strong ($|B_z|>1000G$) fields 
separately.

By analyzing this dataset we find that:
1. For weak magnetic fields, the signs of both
$\alpha$ and current helicity follow the established 
hemispheric rule except during the Year 2004.
The magnitudes of their weighted averages show a weak tendency
of decreasing with the development of solar cycle.
2. For strong magnetic fields, both $\alpha$ and
current helicity show a helicity sign opposite to that
of weak fields.

Our results seem to be consistent 
with the theoretical prediction that solar dynamo would produce 
opposite helicity signs in the mean field and in the fluctuations 
as well as the theoretical calculations of helicity production rate 
by differential rotation. However, as pointed out by the referee, 
some previous studies (Longcope et al. 1998,1999; Chae 2001) have 
suggested that neither the interface dynamo nor the differential 
rotation would generate sufficient amount of helicity (twist). 
So our observation with its interesting implications advocates 
further investigations, both observationally and theoretically.


\acknowledgements

I thank Mitchell Berger for providing his calculation 
data used in the letter.
I also thank the anonymous referee for helpful comments and
suggestions.
This work was supported by the One-Hundred-Talent Program
of the Chinese Academy of Sciences, the Chinese National Science
Foundation under Grant 10373016, and the U.S. National Science
Foundation under Grant ATM-0548060.



\begin{figure}
\epsscale{.85}
\plotone{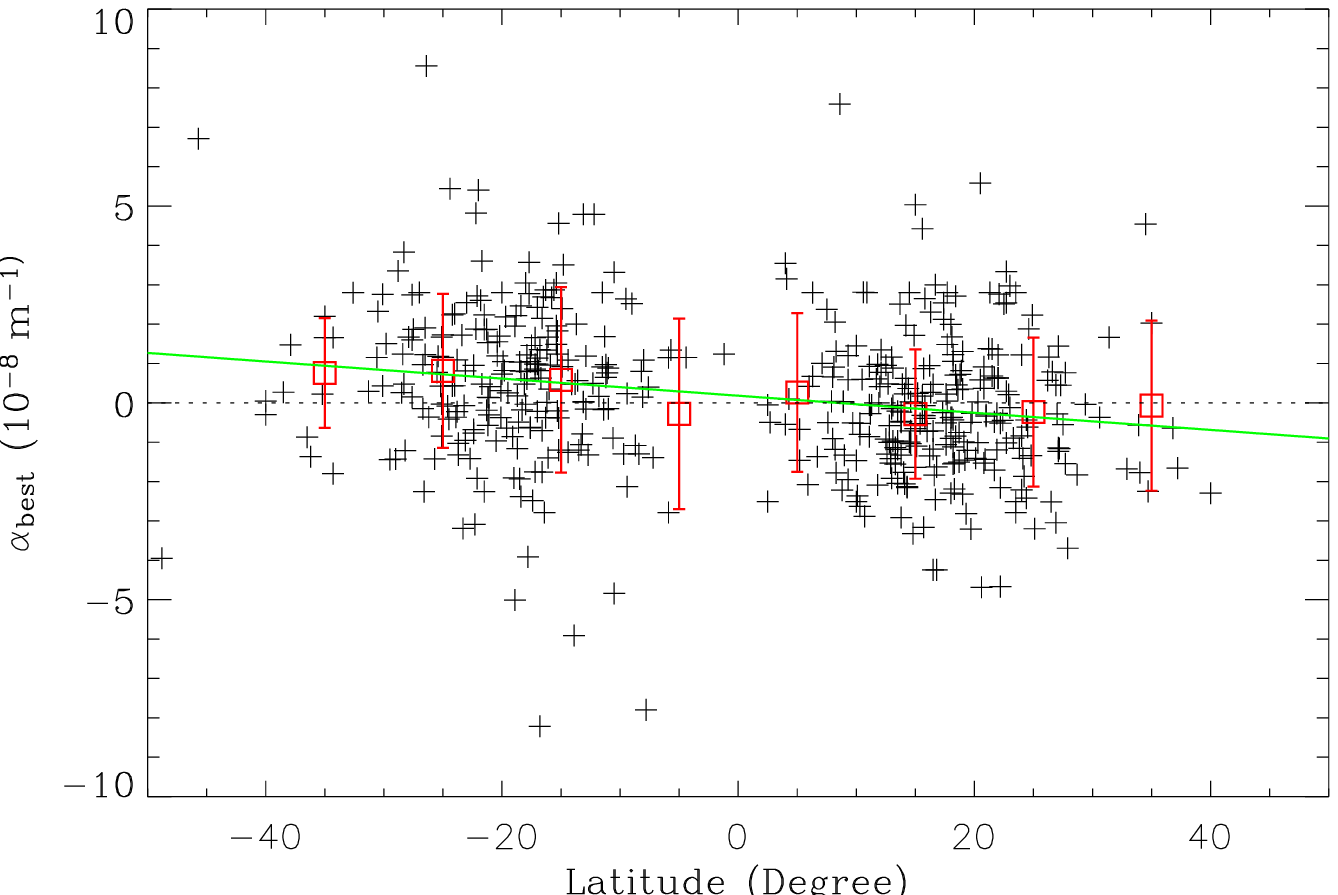}
\caption{Latitudinal profile of $\alpha_{best}$ for the 391
active regions observed by Huairou magnetograph between
1997 July and 2000 September.}
\end{figure}

\clearpage

\begin{figure}
\epsscale{.80}
\plotone{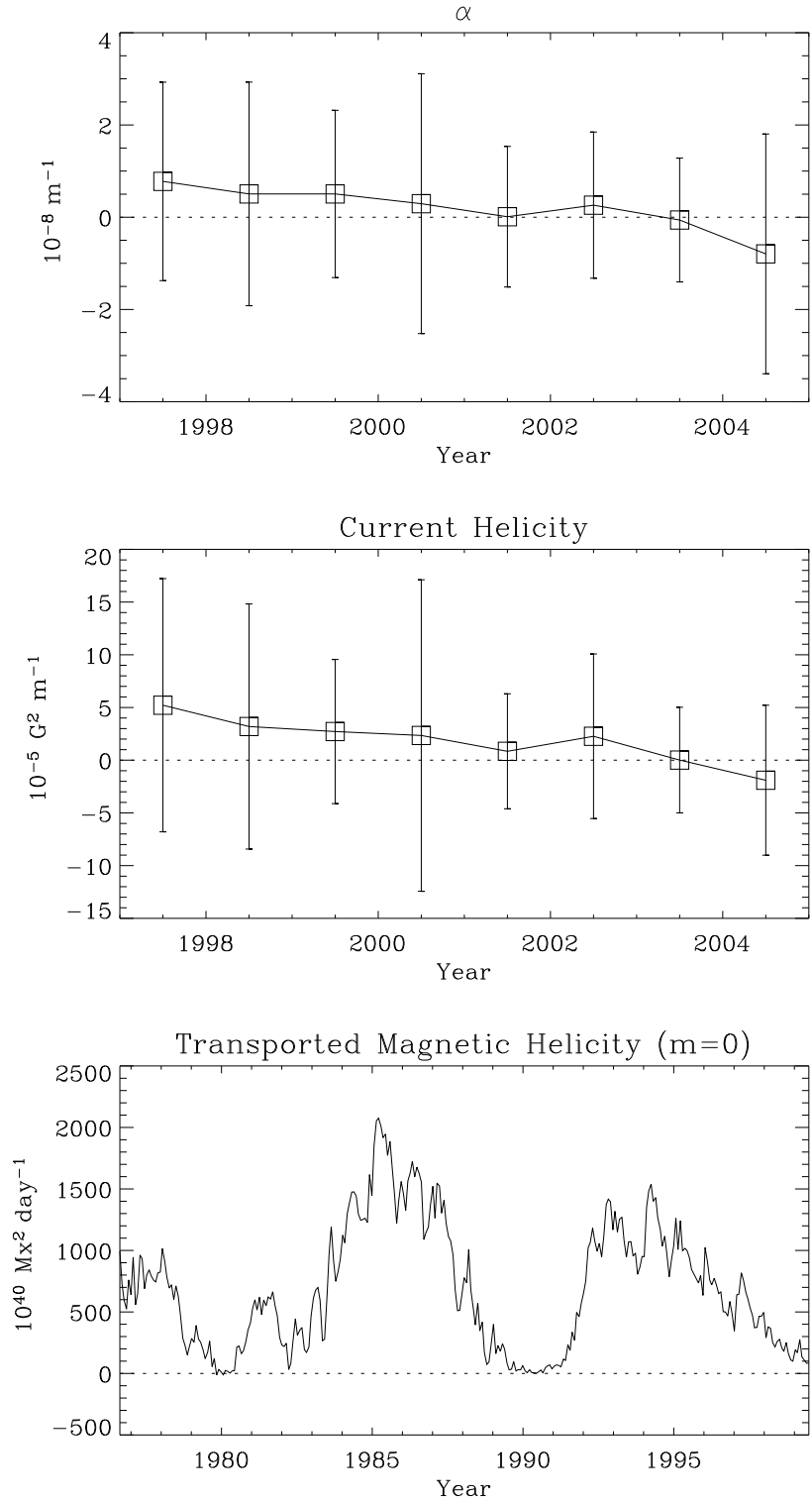}
\caption{Solar cycle variations of weighted averages
of $\alpha$ (top panel) and current helicity
(middle panel) for weak fields ($100G<|B_z|<500G$).
Bottom panel: Calculated transfer rate of $m=0$ mode helicity,
created by differential rotation in the interior, into the 
southern hemisphere. Adopted from Berger \& Ruzmaikin (2000).}
\end{figure}

\clearpage

\begin{figure}
\epsscale{.80}
\plotone{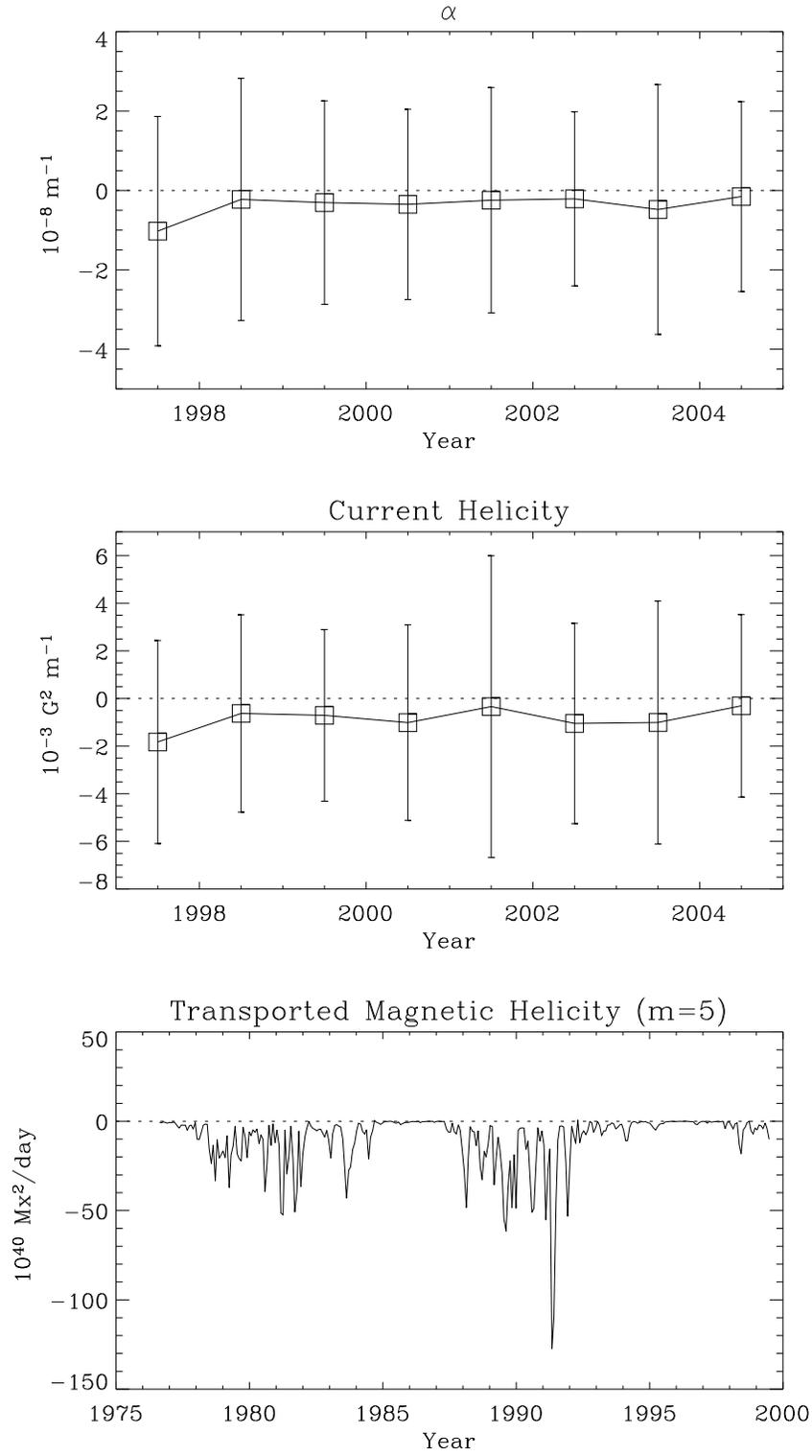}
\caption{Top and middle panels: Same as in Figure 2
but for strong fields ($|B_z|>1000G$). Bottom panel:
Same as in Figure 2 but for $m=5$ mode.}
\end{figure}


\end{document}